\documentclass[aps,prl,reprint,superscriptaddress,amsmath,amssymb,floatfix,showpacs]{revtex4-1}
\usepackage{graphicx}
\usepackage{subfigure}
\usepackage{color}
\usepackage{epstopdf}

\begin{document}

\title{Orbital-selective superconductivity, gap anisotropy and spin resonance excitations
in a multiorbital $t$-$J_1$-$J_2$ model for iron pnictides}

\author{Rong Yu}
\affiliation{Department of Physics \& Astronomy, Rice University, Houston, Texas 77005}
\author{Jian-Xin Zhu}
\affiliation{Theoretical Division, Los Alamos National Laboratory,
Los Alamos, New Mexico 87545}
\author{Qimiao Si}
\affiliation{Department of Physics \& Astronomy, Rice University, Houston, Texas 77005}

\begin{abstract}
We study the orbital-dependent superconducting pairing in a five-orbital $t$-$J_1$-$J_2$
model for iron pnictides.
Depending on the orbital selectivity of electron correlations and
the orbital characters along the Fermi surface,
the superconducting gap in an $A_{1g}$ pairing state may exhibit anisotropy.
This anisotropy varies with the degree of $J_1$-$J_2$ magnetic  frustration.
We have also calculated the dynamical spin susceptibility in the superconducting state.
The frequency dependence of the susceptibility at the antiferromagnetic wavevector
$(\pi,0)$ shows a resonance,
whose width is enhanced by the orbital dependence of the superconducting gap;
when the latter is sufficiently strong, the resonance peak may be split into two.
We discuss the implications of our results on the recent
angle-resolved photoemission and neutron-scattering measurements
in several superconducting iron pnictides.
\end{abstract}

\pacs{71.30.+h, 74.70.Xa, 71.10.Hf, 71.27.+a}

\maketitle

{\it Introduction.~} The mechanism and symmetry of the superconducting pairing is a central issue
for iron-based superconductors~\cite{Kamihara_FeAs,Zhao_Sm1111_CPL08}. In most of these materials,
superconductivity appears when electron or hole doping is introduced into
the antiferromagnetic parent compounds~\cite{Cruz}.
Two theoretical approaches have been proposed to address the link between the magnetism
and superconductivity.
In the weak-coupling limit, both properties arise
from the nesting between electron and hole Fermi pockets ~\cite{Graser09}.
In the strong-coupling approach, on the other hand, the superconductivity and
antiferromagnetism
are driven by the short-range exchange interactions among the correlation-induced
quasi-local moments
~\cite{SiAbrahams08,Yildirim08,MaXiang08,Fang08,Dai09}.
The superconducting gap function reflects the
short-range electron pairs, and is explicitly orbital dependent.
Among the evidences~\cite{YuGoswami11,DHLee13}
for the strong-coupling approach is the fact
that both the alkaline iron selenides $K_{1-x}$Fe$_{2-y}$Se$_2$
~\cite{Zhang_Feng,Qian_Ding,Mou}
and the single-layer FeSe ~\cite{DLiu12}
lack the Fermi surface nesting (in the absence of hole Fermi pockets) and
yet still display high-$T_c$
superconductivity.

A recent development in the strong-coupling approach to the iron-based superconductors is the
proposal that the proximity to the Mott transition is orbital-dependent
~\cite{YuSi11,YinKotliar11,YuSi13,YinKotliar12,Bascones12,YuZhuSi13}.
Experimental evidence for such orbital selectivity has
already emerged~\cite{Yi13,Lee12}.
It is therefore natural to ask
how such orbital-dependent effects of the electron correlations influence
the nature of the superconducting state.

The pairing symmetry of the iron-based superconductors
has been studied via various experimental techniques.
Angle resolved photoemission spectrum (ARPES) measurements find the superconducting gap to be
nodeless and isotropic at both the hole and electron pockets in a number
of materials~\cite{Ding08,Kondo08,Liu11,Xu11,RichardDing11}. Neutron scattering measurements
on these compounds observe a clear spin resonance mode
in the superconducting state~\cite{Christianson08,ZhangDai11,Lumsden09,Chi09}.
This is consistent with the pairing order parameter changing sign
between the hole pockets near the Brillouin zone (BZ) center and the electron pockets near
the zone boundary,
which arises within both weak-coupling~\cite{Kuroki08,Mazin08}
and strong-coupling approaches~\cite{Seo08,Goswami10}.

Recently, experiments have identified an anisotropy
of the superconducting gap along the Fermi pockets in several iron pnictide materials~\cite{GeFeng13,UmezawaWang12,AllanDavis12,Hackl12}. These results hold the promise
to shed new light on the understanding of the pairing mechanism.
In particular, high resolution ARPES~\cite{GeFeng13} has revealed that
the superconducting gap is anisotropic
along the electron Fermi pockets in the underdoped Na(Fe$_{1-x}$,Co$_x$)As;
it becomes isotropic when the system reaches the overdoped regime.
Because the orbital character of the electronic  states varies along the electron Fermi pocket,
this observation points to  the possibility of the orbital dependent nature of superconductivity.

In this Letter, we study the superconducting pairing in a five-orbital $t$-$J_1$-$J_2$ model for iron pnictides.
We show that the orbital-dependent effects of electron correlations generally give rise to
orbital-selective superconducting gaps. In particular, we emphasize two gaps that are
respectively associated with the $xz/yz$ and $xy$ orbitals, which
have the same $A_{1g}$ symmetry but
different pairing amplitudes. We show how this orbital-selective pairing naturally
leads to a gap anisotropy, and
discuss the implication of the results for the ARPES and neutron spectra of several superconducting
iron pnictides.

{\it Model and method.~} We consider a five-orbital $t$-$J_1$-$J_2$ model. The Hamiltonian~\cite{Goswami10,Seo08,SiAbrahams08} reads as
\begin{eqnarray}
H&=&-\sum_{i<j,\alpha,\beta,s} \left( t_{ij}^{\alpha \beta}
c^{\dagger}_{i\alpha s}c_{j\beta s}+h.c. \right)
+\sum_{i,\alpha} (\epsilon_\alpha-\mu) n_{i\alpha} \nonumber \\
&&+\sum_{\langle ij\rangle,\alpha,\beta} J_{1}^{\alpha \beta}
\left(\mathbf{S}_{i\alpha}\cdot \mathbf{S}_{j\beta}-\frac{1}{4}n_{i\alpha} n_{j\beta}\right)\nonumber \\
&&+\sum_{\langle \langle ij\rangle \rangle,\alpha,\beta} J_{2}^{\alpha \beta}
\left(\mathbf{S}_{i\alpha}\cdot \mathbf{S}_{j\beta}-\frac{1}{4}n_{i\alpha} n_{j\beta}\right),
\end{eqnarray}
where $c^{\dagger}_{i\alpha s}$
creates an electron at site $i$, in orbital $\alpha$ and spin
projection $s$; $\mu$ is the chemical potential to fix the total electron number $n$.
The orbital index $\alpha=1,2,3,4,5$ respectively correspond to the five Fe $3\mathrm{d}$
orbitals $xz$, $yz$, $x^2-y^2$, $xy$, and $3z^2-r^2$.
The tight-binding parameters $t_{ij}^{\alpha \beta}$ and $\epsilon_\alpha$
respectively refer to the hopping matrix and the onsite potential that reflects the crystal level splitting.
For definiteness, we consider the case of NaFeAs,
and obtain the tight-binding parameters by fitting its
LDA bandstructure
~\cite{Suppl}.
The nearest-neighbor (n.n., $\langle ij\rangle$)
and next-nearest-neighbor (n.n.n., $\langle \langle ij \rangle \rangle$)
exchange couplings are respectively denoted by $J_{1}^{\alpha \beta}$ and $J_{2}^{\alpha \beta}$. The spin
 operator is
$\mathbf{S}_{i\alpha}=\frac{1}{2}\sum_{s,s^{'}}c^{\dagger}_{i\alpha s}
\boldsymbol{\sigma}_{ss^{'}}c_{i\alpha s^{'}}$
and  the density operator
$n_{i \alpha}=\sum_{s}c^{\dagger}_{i\alpha s}c_{i\alpha s}$,
where $\boldsymbol{\sigma}$ represents the
Pauli matrices.
The double-occupancy prohibiting constraint from the fermion is implicitly
incorporated by the renormalization of the band structure~\cite{Goswami10,YuGoswami11}.

We study the superconducting pairing in the above $t$-$J_1$-$J_2$ model by
decomposing the exchange interactions in the
spin singlet pairing channels.
The gliding reflection symmetry of the Fe-As lattice allows us to consider
the pairing channels with the choice of the one-Fe unit cell.
For simplicity, we assume
$J_{1(2)}^{\alpha \beta}=J_{1(2)}\delta_{\alpha \beta}$ (and take $J_2$
to be the energy unit); correspondingly, we consider
intraorbital pairing.
There are 20 different pairing channels, each with an amplitude and a phase,
which are self-consistently determined.

We also calculate the dynamical spin susceptibility in the superconducting state.
At wavevector $\mathbf{q}$ and Matsubara frequency $\omega_n$ the
spin susceptibility $\chi(\mathbf{q},i\omega_n)=\sum_{\alpha\beta} \chi_{\alpha,\beta}(\mathbf{q},i\omega_n)$,
where  $\chi_{\alpha,\beta}(\mathbf{q},i\omega_n) = \sum_\gamma \left(\mathbf{I}+J(\mathbf{q})\boldsymbol{\chi}^0
(\mathbf{q},i\omega_n)\right)^{-1}_{\alpha\gamma} \chi^0_{\gamma,\beta}(\mathbf{q},i\omega_n)$,
and $\chi^0_{\alpha,\beta}(\mathbf{q},i\omega_n) = \int^{1/T}_0 d\tau e^{i\omega_n\tau}\langle\mathcal{T}_\tau
[S^-_{\mathbf{q}\alpha}(\tau) S^+_{-\mathbf{q}\beta}(0)]\rangle$. Here $J(\mathbf{q})=\frac{J_1}{2}
(\cos q_x+\cos q_y)+J_2\cos q_x \cos q_y$, $S^{\pm}_{\mathbf{q}\alpha}
= \frac{1}{\sqrt{N}}\sum_i e^{i\mathbf{q}\cdot\mathbf{r}_i} S^{\pm}_{i\alpha}$, and $\langle\ldots\rangle$
refers to the expectation value
with respect to the effective Hamiltonian.
The susceptibility at real frequency $\omega$ is then obtained by an
analytical continuation $i\omega_n\rightarrow \omega+i0^+$.

\begin{figure}[t!]
\centering\includegraphics[
bb = 0 25 588 390,clip,
width=85mm]{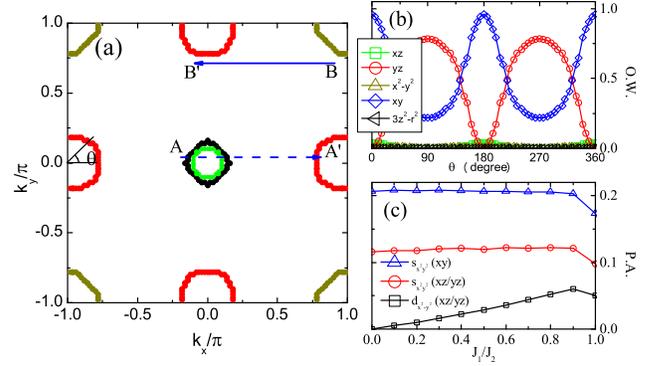}
\caption{(Color online)
(a): Fermi surface in the one-Fe Brillouin zone of the five-orbital tight-binding model at electron
doping $x=0.02$. Angle $\theta$ parameterizes the Fermi surface pockets in the momentum space.
The arrows indicate the dominant scattering processes contributing to the spin resonance
peaks shown in Fig.~\ref{fig:3}(a). (b): The orbital characters along the electron pocket near $(\pi,0)$.
Here O.W. denotes orbital weight. (c): Evolution of the leading pairing channels in the $t$-$J_1$-$J_2$ model with $J_1/J_2$.
They all have the $A_{1g}$ symmetry. Here P.A. denotes pairing amplitude.
}
\label{fig:1}
\end{figure}

{\it Multiorbital nature of the Fermi surface and orbital-selective pairing.~} The Fermi surface in the 1-Fe BZ
for the tight-binding model at electron doping $x=0.02$ ($x=n-6$) is shown in Fig.~\ref{fig:1}(a).
The Fermi surface contains multiple sheets with different orbital characters. The two hole pockets near $(0,0)$
are dominated by the degenerate $xz/yz$ orbitals; the hole pocket near $(\pi,\pi)$ has almost
completely $xy$ orbital character. The electron pocket near $(\pi,0)$ [or $(0,\pi)$] displays a hybridized $xy$
and $yz$ ($xz$) orbital character [Fig.~\ref{fig:1}(b)]. The pairing amplitudes are also orbital dependent.
For $J_1/J_2\lesssim1$, the dominant pairing channel is $s_{x^2y^2}$ with
an $A_{1g}$ symmetry (Fig.~\ref{fig:1}(c)).
The amplitude of this pairing channel in the $xy$ orbital is larger than that in the $xz/yz$ orbital.
Due to the double degeneracy of the $xz$ and $yz$ orbitals, a $d_{x^2-y^2}$ wave pairing channel
can also have an $A_{1g}$ symmetry. It serves as a subdominant pairing channel whose amplitude
increases with $J_1/J_2$.
The existence of orbital-selective multiple energy scales in pairing is a consequence of the orbital dependent
electron correlation effects in the multiorbital model, with the $xy$ orbital typically exhibiting strong
correlation effects ~\cite{YinKotliar11,YuSi11,YuSi13}. Correspondingly, the $xy$ orbital has a sizeable ratio of $J$
to the renormalized bandwidth, which in turn yields a sizeable pairing amplitude
~\cite{YuGoswami11}.

\begin{figure}[t!]
\centering\includegraphics[
width=85mm]{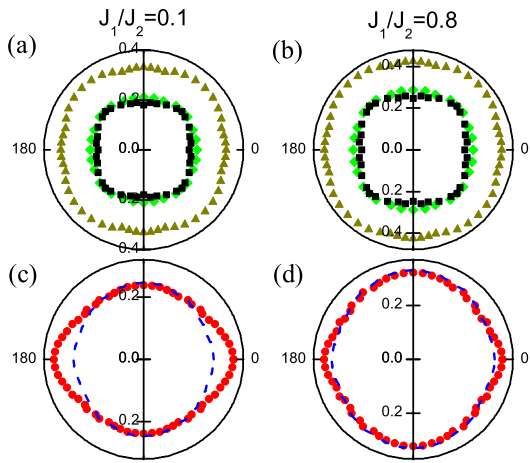}
\caption{(Color online)
Angular dependence of the excitation gaps of BCS quasiparticles along the Fermi pockets
in the $t$-$J_1$-$J_2$ model at $J_1/J_2=0.1$ (in (a) and (c)) and $J_1/J_2=0.8$ (in (b) and (d)),
respectively. In (a) and (b), green diamonds and black squares refer to the gaps along the inner and
outer hole pockets
near $(0,0)$; brown triangles refer to the gap along the hole pockets near $(\pi,\pi)$.
In (c) and (d), red circles refer to the gap along the electron pocket near $(\pi,0)$.
The blue dashed line is a fit to the single parameter gap function $\Delta_0\cos k_x \cos k_y$.
The deviation from this fit implies a multi-gap structure of the multiorbital model (see text).
}
\label{fig:2}
\end{figure}

{\it Anisotropic superconducting gap.~}
We now turn to how the orbital-selective pairing amplitudes and the orbital character of the Fermi surface
affect the momentum distribution of the superconducting gaps by inducing gap anisotropy,
and how the gap amplitudes and the corresponding anisotropy can be tuned by the degree of magnetic
frustration of the system. We discuss and compare the results in the five-orbital $t-J_1-J_2$
model by taking $J_1/J_2=0.1$ and $J_1/J_2=0.8$, for illustrative purpose.
For $J_1/J_2=0.1$, the superconducting
gaps are dominated by the $s_{x^2y^2}$ $A_{1g}$ pairing channel.
The amplitudes of this pairing channel in the $xy$
and $xz/yz$ orbitals are significantly different, resulting in two characteristic gaps $\Delta_{xy}\neq\Delta_{xz/yz}$.
The excitation gap of the
quasiparticles along each hole pocket is only associated with one of them
[see Supplementary Material ~\cite{Suppl}, Figs.~S3 (a) and (b) ],
and is isotropic [Fig.~\ref{fig:2}(a)] since the dominant orbital character of a hole pocket is uniform:
$xy$ for the pocket near $(\pi,\pi)$ and $xz/yz$ for the pocket near $(0,0)$.
 On the other hand, the gaps along the electron pockets are strongly anisotropic [Fig.~\ref{fig:2}(c)].
 This is because the electron pocket has a hybridized $xy$ and $xz/yz$ orbital character, and the size of the gap
 at a particular wavevector depends on the dominant orbital character at that point. The gap anisotropy
 reflects these two characteristic superconducting gaps $\Delta_{xy}\neq\Delta_{xz/yz}$: as shown in Fig.~\ref{fig:2}(c),
 the gap cannot be fitted by a single gap function $\Delta_0\cos k_x \cos k_y$ though the dominant pairing
 channel is $s_{x^2y^2}$ $A_{1g}$. Interestingly, the gap anisotropy reduces with increasing
 $J_1/J_2$,
 and an essentially isotropic gap along the electron pocket is recovered at $J_1/J_2=0.8$ (Fig.~\ref{fig:2}(d)).
 To understand this, note that the pairing amplitude of the subdominant $d_{x^2-y^2}$ $A_{1g}$ channel
 in the $xz/yz$ orbital increases with $J_1/J_2$. With the contribution from this subdominant channel, the overall
 gap in the $xz/yz$ orbital $\Delta_{xz/yz}\approx\Delta_{xy}$.
 This then leads to an essentially isotropic gap along the electron pockets.

\begin{figure}[t!]
\centering\includegraphics[
width=85mm]{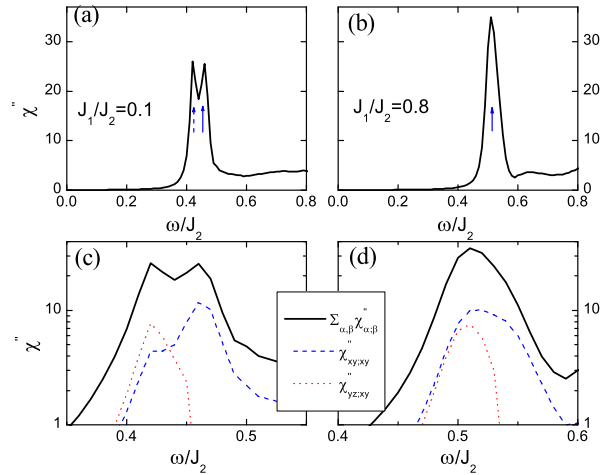}
\caption{(Color online)
Calculated imaginary part of the dynamical spin susceptibility $\chi^{\prime\prime}(\mathbf{q},\omega)$ at wavevector $\mathbf{q}=(\pi,0)$ in the $t$-$J_1$-$J_2$ model for $J_1/J_2=0.1$ (in (a) and (c)) and $J_1/J_2=0.8$ (in (b) and (d)), respectively. Also shown in (c) and (d): the orbital resolved dominant components of the susceptibility.
}
\label{fig:3}
\end{figure}

{\it Spin resonance excitation.~}
The spin excitations in the superconducting state are also affected by the orbital-selective pairing.
We have calculated the dynamical spin susceptibility in the superconducting state for
the $J_1/J_2=0.1$ and $J_1/J_2=0.8$ cases discussed above.
The imaginary part of the susceptibility
$\chi^{\prime\prime}(\mathbf{q},\omega)$
at the antiferromagnetic wavevector $\mathbf{q}=(\pi,0)$
 exhibits two resonance peaks  in the frequency dependence
 for $J_1/J_2=0.1$ (Fig.~\ref{fig:3}(a)).
Our detailed analysis finds that the double peak structure of $\chi^{\prime\prime}(\mathbf{q},\omega)$ arises
from the different scattering processes that connect two regimes near the electron and hole pockets,
as indicated by the arrows
in Fig.~\ref{fig:1}(a). By appearing in the coherence factor of the expression
of $\chi^{\prime\prime}(\mathbf{q},\omega)$,
the different orbital characters of the quasiparticle dispersion put a strong constraint to the scattering processes
such that the spin response is enhanced only in certain regimes of the BZ, where the orbital characters
of the associated hole and electron bands
  are compatible. For example, the dominant contribution to the lower frequency resonance peak at $\omega_L$
(see Fig.~\ref{fig:3}(c)) is from a scattering between the $yz$ orbital in regime A and the $xy$
orbital in regime A$^{\prime}$,
as indicated by the dashed arrow in Fig.~\ref{fig:1}(a). The higher frequency resonance peak at $\omega_H$,
on the other hand, is mainly associated with a scattering within the $xy$ orbital between regimes B and
B$^{\prime}$
(see Fig.~\ref{fig:3}(c) and Fig.~\ref{fig:1}(a)).
As a rough estimate, the resonance frequency $\omega\lesssim E_h+E_e$,
where $E_h$ and $E_e$  are respectively the excitation gaps of the
corresponding hole- and electron-like quasiparticles,
\emph{i.e.} $h=\mathrm{A},\mathrm{B}$, and $e=\mathrm{A}^{\prime},\mathrm{B}^{\prime}$.
Given the similar orbital character and the proximity to the equivalent
points along the Fermi surface, $E_{A^{\prime}}\approx E_{B^{\prime}}$.
 But the different orbital characters make $E_A\neq E_B$ for $J_1/J_2=0.1$.
 (Note that as a combined effect of the momentum dependence of the gap function
 and the multiorbital nature, $E_B$ is smaller than the gap along the
 nearby hole
 Fermi
 pocket~\cite{Suppl}.) Therefore, $\omega_L\neq\omega_H$;
 when this difference is sufficiently large, two resonances
appear in the frequency-dependent spectrum.
 As $J_1/J_2$ increases, both $E_A$ and $E_B$ increase. But due to the subdominant $d_{x^2-y^2}$ channel
 in the $xz/yz$ orbital, $E_A$ increases faster, and $E_A\approx E_B$ for $J_1/J_2=0.8$.
 We thus obtain a single resonance peak at $\omega_L\approx\omega_H$, as shown in Fig.~\ref{fig:3}(b) and (d));
 the multi-orbital effect is then reflected in the broadening of the peak.

 {\it Discussions.~}
Our results elucidate how the orbital selectivity of electron correlations influences the superconductivity.
We show that the orbital-selective pairing gives rise to gap anisotropy along a Fermi surface with hybridized orbital
characters.
By promoting the subdominant $d_{x^2-y^2}$ $A_{1g}$ pairing channel, the magnetic frustration may compete
with the orbital selectivity, and tune the gap anisotropy.
Our results are particularly pertinent
to the anisotropic superconducting gap along the electron pockets in the underdoped
Na(Fe$_{1-x}$,Co$_x$)As observed in recent ARPES measurements~\cite{GeFeng13}.
We are also able to understand the evolution from the anisotropic to isotropic gap with increasing electron doping.
For illustrating purposes, in the model calculation of this paper,
we fix the electron doping and show the evolution of gap anisotropy by tuning $J_1/J_2$.
But the gap anisotropy as a consequence of the orbital selectivity is a general result.
In a more realistic model, the exchange couplings would be orbital dependent.
Whether an anisotropic gap shows up depends on how different the ratio of exchange coupling
to the renormalized electronic-bandwidth,  $J_{1(2)}^{\alpha}/D^{\alpha}$, is
among the different orbitals, and this ratio will be tuned by the strength of electron correlations.

Theoretically, the coexistence of antiferromagnetic order and superconductivity
in the underdoped regime may also
lead to an anisotropic superconducting gap along the reconstructed Fermi
surface~\cite{MaitiChubukov12}.
In this scenario, the Fermi surface is reconstructed for both the electron and hole pockets.
But experimentally,
the gap anisotropy was only observed along electron pockets. Moreover, the Fermi surface reconstruction
for the gap anisotropy was not observed in ARPES~\cite{GeFeng13}. It therefore is unlikely that
the observed anisotropic gap is primarily driven by the coexistence of superconductivity with antiferromagnetic order.

Raman scattering has also implicated an anisotropic gap
along the electron pocket in the nearly optimally hole-doped
BaFe$_2$As$_2$.~\cite{Hackl12}  Because hole doping tends to increase
the orbital selectivity
of electron correlations~\cite{deMedici12}, it is natural to propose that the mechanism
advanced here underlies this
experimental observation as well.

We have also shown that the frequency dependence of the dynamical spin susceptibility
at $(\pi,0)$ displays a resonance, whose width is enhanced by the orbital-dependence
of the superconducting gap. When the latter is sufficiently strong,
the resonance peak may be split into two. We propose that this mechanism underlies
the recent neutron-scattering
observation of double spin resonances in the electron underdoped NaFeAs system~\cite{ZhangDai13}.
The double-resonance feature we have discussed is very different from the one reported
in Ref.~\cite{DasBalatsky11}.
In that case, the wavevectors where the resonances take place are sensitive to the
Fermi surface geometry,
and are necessarily different for the two resonances; one is at $\mathbf{q}=(\pi,0)$, while the other
is at an incommensurate $\mathbf{q}$. In our case, both resonances take place at the same
wavevector
$\mathbf{q}=(\pi,0)$;
this wavevector is determined by the $\mathbf{q}$-dependence of $J(\mathbf{q})$.

Finally, the degree of electron correlations remains a central issue in the iron-based superconductors.
This issue is typically probed in the normal state, through the bad-metal phenomenology in
the optical spectrum \cite{Qazilbash09} or the orbital selectivity in the ARPES spectrum \cite{Yi13,Lee12}.
Our theoretical results here suggest that this issue can also be fairly directly probed through the
orbital selectivity of the gap function in the superconducting state. ARPES studies along this direction
are already quite realistic~\cite{Shimojima11,Malaeb12},
and we anticipate that considerable new insights will be derived through
further studies along this direction.

{\it Conclusions.~}
Our calculation on the superconducting pairing in a five-orbital $t$-$J_1$-J$_2$ model for iron pnictides reveals
an orbital-selective gap structure due to the strong electron correlation effects.
While both gaps have the $s_{x^2y^2}$ $A_{1g}$ symmetry, the different orbital character gives
rise to gap anisotropy along the electron pockets. The orbital selective pairing leads to a broadened
neutron
resonance at the antiferromagnetic ordering wavevector $\mathbf{q}=(\pi,0)$ in the superconducting state.
This resonance may even be split into two peaks.
Our results have important implications
for  the ARPES and neutron measurements on the electron underdoped NaFeAs,
as well as the Raman scattering results on the hole doped BaFe$_2$As$_2$.
More generally, our results point to new ways of probing electron correlation effects of the iron pnictides
through the
single-particle and spin responses in their superconducting state.

{\it Acknowledgements.~} We thank P.\ Dai, D.\ H.\ Lu, and C.\ L.\ Zhang for useful discussions.
This work has been supported by the NSF, the Robert A.\ Welch Foundation
Grant No.\ C-1411,
and the Alexander von Humboldt Foundation. One of us (Q. S.)
acknowledges the hospitality of the Aspen Center for Physics (NSF Grant No. 1066293),
the Institute of Physics of Chinese Academy of Sciences,
and the Karlsruhe Institute of Technology.

\setcounter{figure}{0}
\makeatletter
\renewcommand{\thefigure}{S\@arabic\c@figure}

\setcounter{table}{0}
\makeatletter
\renewcommand{\thetable}{S\@arabic\c@table}

\onecolumngrid

\section*{Supplementary Material}
\subsection{Tight-binding parameterization}
To obtain the tight-binding
parameters, we perform LDA calculations for NaFeAs, and fit the LDA bandstructure to the
tight-binding Hamiltonian.
We use the form of the
five-orbital tight-binding Hamiltonian given in Ref.~\cite{Graser09}.
The tight-binding parameters so derived are listed in Table~\ref{tab:1}.

\begin{figure}[h!]
\centering\includegraphics[
width=100mm]{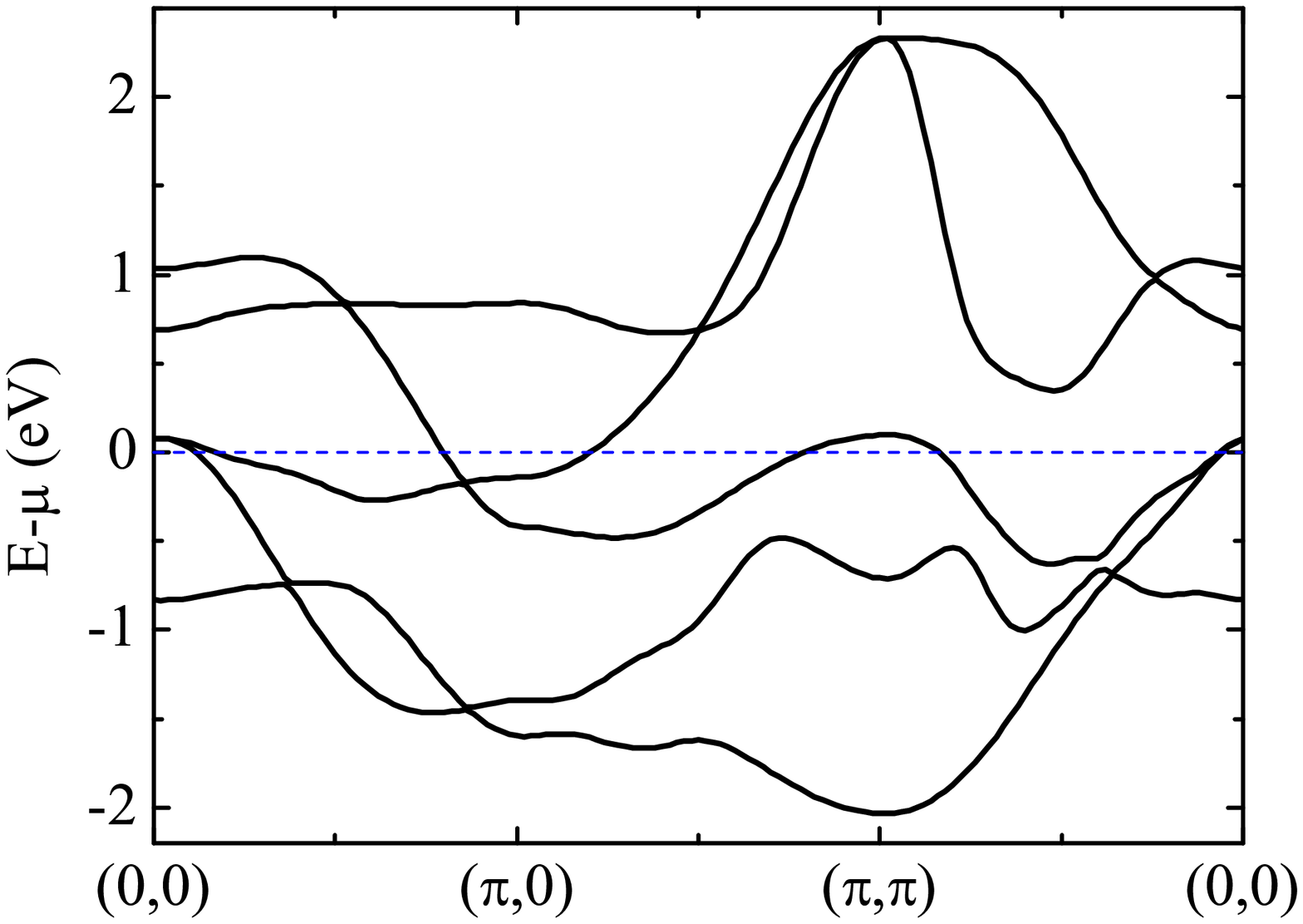}
\caption{(Color online)
Bandstructure of the five-orbital tight-binding model at $n=6.02$ along high-symmetry directions of the
one-Fe Brillouin zone.
}
\label{fig:S1}
\end{figure}

Fig.~\ref{fig:S1} shows the bandstructure of the five-orbital tight-binding model for electron density $n=6.02$,
corresponding to $x=0.02$ electron doping. The corresponding Fermi surface is shown in Fig.~1 of the main text.
The Fermi surface consists of three hole pockets and two electron pockets.
They have very different orbital compositions.
We show the orbital weights of the hole and electron pockets in Fig.~\ref{fig:S2}(b) and Fig.~\ref{fig:S3}.

\subsection{Momentum distribution of the excitation gap of the quasiparticles}

\begin{figure}[h!]
\centering\includegraphics[
bb = 0 0 490 450,clip,
width=75mm]{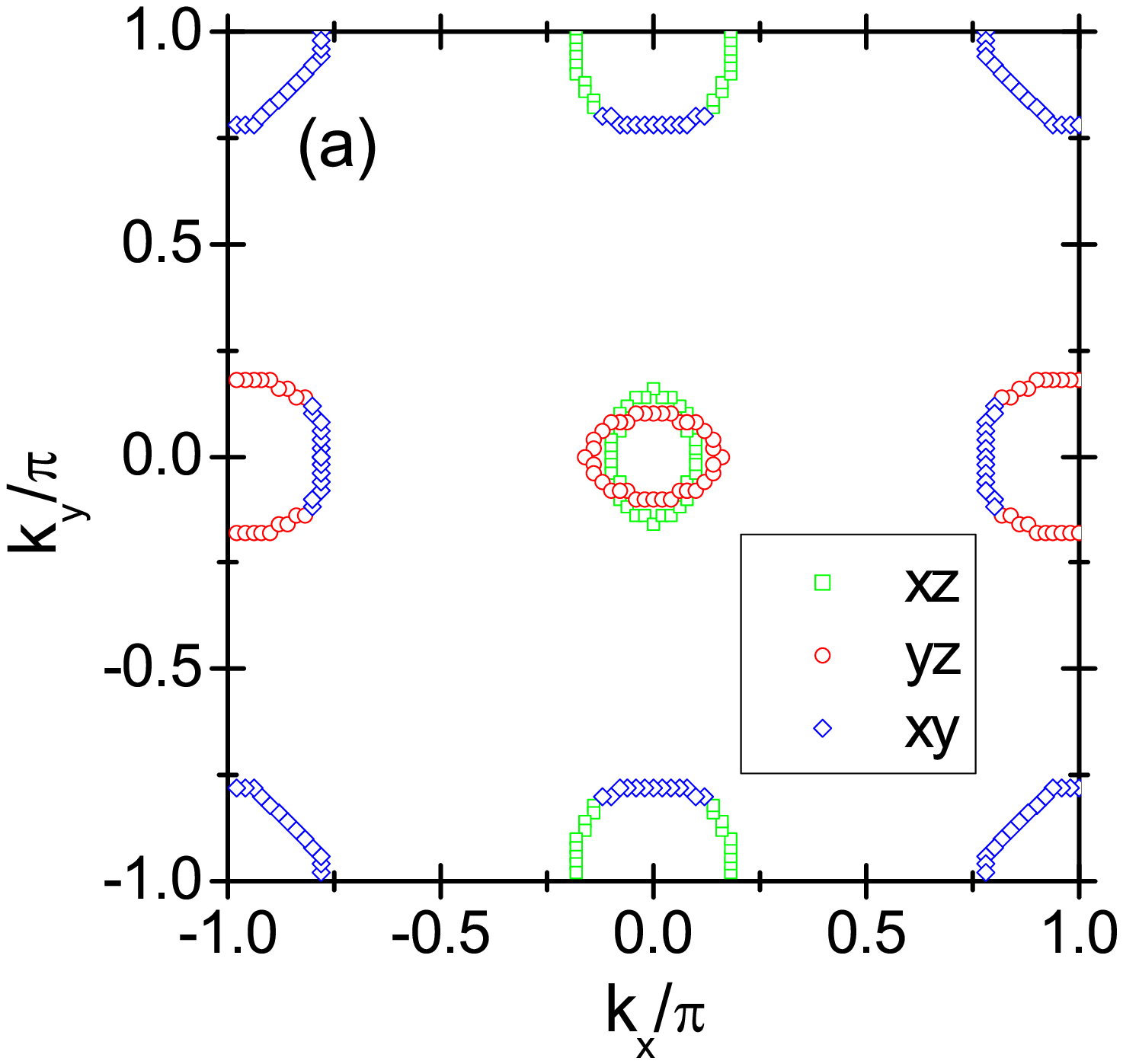}
\centering\includegraphics[
bb = 90 60 280 220,clip,
width=80mm]{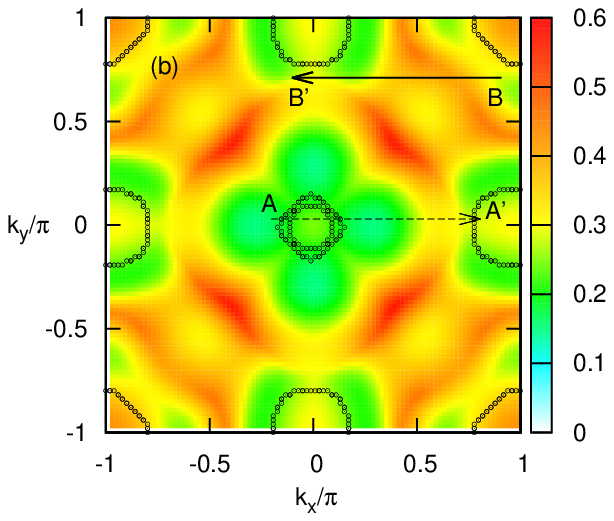}
\caption{(Color online)
(a): Fermi surface of the five-orbital tight-binding model at $n=6.02$. Different symbols represent the dominant
orbital characters of the pockets.
(b): Momentum distribution of the excitation gap of the quasiparticles in the five-orbital $t$-$J_1$-$J_2$
model for $n=6.02$ and $J_1/J_2=0.1$. The black circles show the Fermi surface of the tight-binding model
at the same filling. The arrows illustrate the scattering processes that contribute largest to the spin
susceptibility in the superconducting state.
}
\label{fig:S2}
\end{figure}

\begin{figure}[h!]
\centering\includegraphics[
width=150mm]{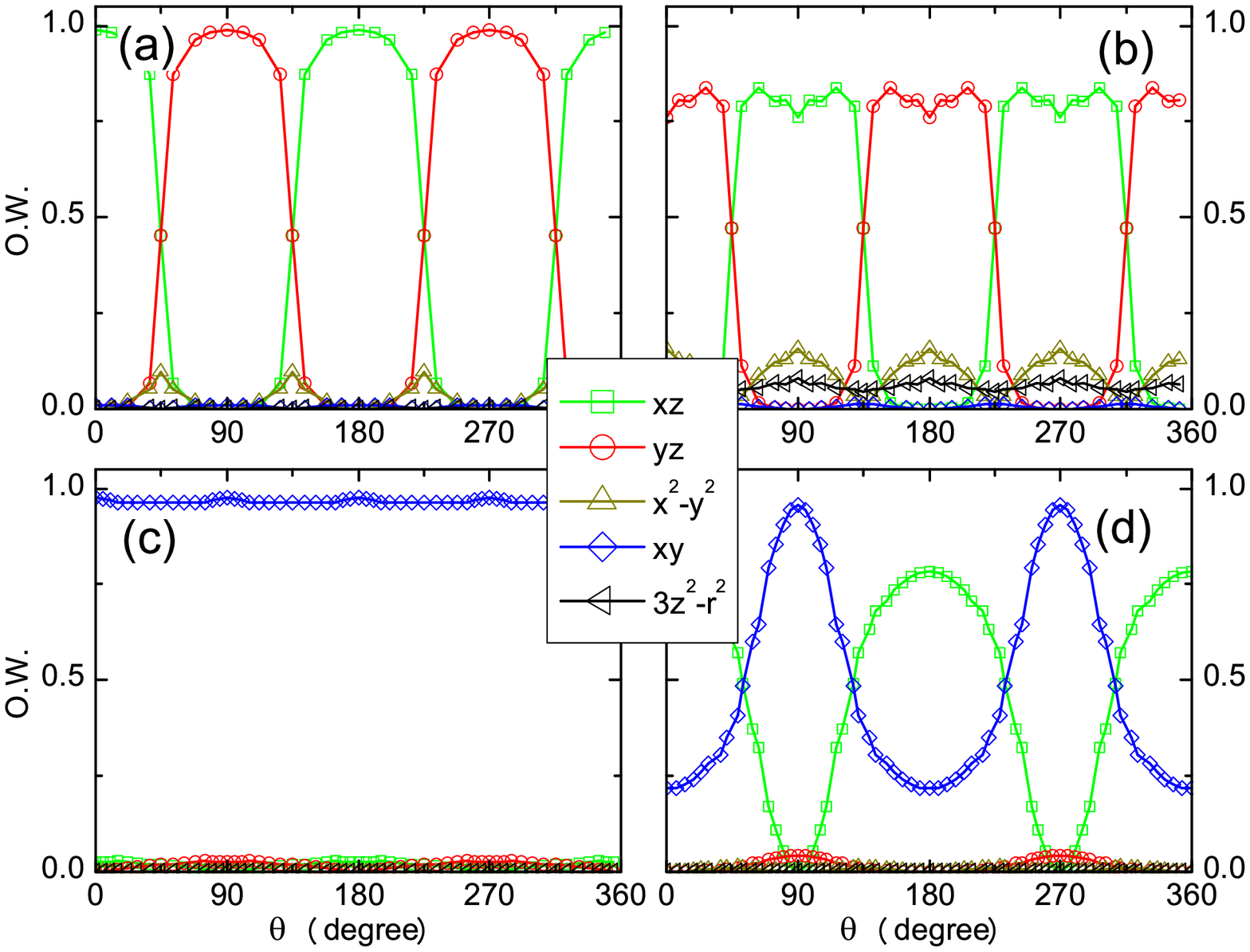}
\caption{(Color online)
Orbital weights along the Fermi surface of the five-orbital tight-binding model at $n=6.02$. (a) and (b): inner and outer hole pockets near $(0,0)$; (c): hole pockets near $(\pi,\pi)$; (d): electron pocket near $(0,\pi)$.
}
\label{fig:S3}
\end{figure}

In the conventional BCS theory for a single-band model with $s$-wave pairing symmetry,
the superconducting gap
$\Delta$ is momentum independent, and the excitation gap for the BCS quasiparticles
is $E(\mathbf{k})=\sqrt{(\xi_{\mathbf{k}}-\mu)^2+\Delta^2}$, where $\xi_{\mathbf{k}}$ and $\mu$
are respectively the dispersion and chemical potential of the tight-binding model.
For the five-orbital $t$-$J_1$-$J_2$ model, considering the multiorbital nature of the model and the
complicated
structure of the superconducting pairing function, the excitation gap has a complicated momentum distribution,
which can only be obtained numerically. We show the momentum distribution of the excitation gap
for $J_1/J_2=0.1$ in Fig.~\ref{fig:S2}. Note that due to the momentum dependent pairing function
and the nonzero inter-orbital hopping, the excitation gap at regime B ($E_B$) is smaller than that
along the hole pocket centered at $(\pi,\pi)$. But still $E_B>E_A$ at $J_1/J_2=0.1$.

\begin{table}
  \centering
\begin{tabular}{cccccccc}
  \hline
  \hline
    & $\alpha=1$ & $\alpha=2$ & $\alpha=3$ & $\alpha=4$ & $\alpha=5$ &   &   \\ \hline
  $\epsilon_\alpha$ & -0.10818 & -0.10818 & -0.40863 & 0.14158 & -0.40471 &  & \\ \hline\hline
  $t^{\alpha\alpha}_\mu$ & $\mu=x$ & $\mu=y$ & $\mu=xy$ & $\mu=xx$ & $\mu=xxy$ & $\mu=xyy$ & $\mu=xxyy$ \\ \hline
  $\alpha=1$ & 0.01398 & -0.42534 & 0.24665 & -0.02238 & -0.00638 & -0.06954 & 0.07281\\ \hline
  $\alpha=3$ & 0.34046 &  & -0.08566 & 0.01052 &  &  &  \\ \hline
  $\alpha=4$ & 0.16907 &  & 0.12337 & 0.00955 & -0.02595 &  & -0.03576 \\ \hline
  $\alpha=5$ & -0.04400 &  &  & -0.04958 & 0.01441 &  & -0.05132 \\ \hline\hline
  $t^{\alpha\beta}_\mu$ & $\mu=x$ & $\mu=xy$ & $\mu=xxy$ & $\mu=xxyy$ &  &  &  \\ \hline
  $\alpha\beta=12$ &  & 0.22625 & -0.06712 & 0.05439 &  &  &  \\ \hline
  $\alpha\beta=13$ & -0.32770 & 0.04340 & 0.03380 &  &  &  &  \\ \hline
  $\alpha\beta=14$ & 0.00011 & -0.10269 & 0.00780 &  &  &  &  \\ \hline
  $\alpha\beta=15$ & -0.04573 & -0.14882 &  & -0.00124 &  &  &  \\ \hline
  $\alpha\beta=34$ &  &  & -0.04511 &  &  &  &  \\ \hline
  $\alpha\beta=35$ & -0.25003 &  & 0.01931 &  &  &  &  \\ \hline
  $\alpha\beta=45$ &  & -0.13024 &  & 0.01023 &  &  &  \\ \hline
  \hline
\end{tabular}
\caption{Tight-binding parameters of the five-orbital model for NaFeAs}\label{tab:1}
\end{table}

\end{document}